\documentclass[preprint,showpacs,preprintnumbers,amsmath,amssymb]{revtex4}

 \usepackage{epsfig}

\begin{document}
\title{ \bf Phase transition in the bounded one-dimensional multitrap system}
\makeatother
\bigskip \bigskip

 \author{  D. Bar$^{a}$}
 \affiliation{ $^a$Department of Physics, Bar Ilan University, Ramat Gan,
Israel.}
 
\bigskip

\begin{abstract}

\noindent
{ \it We have previously   discussed  the diffusion limited problem of the bounded
one-dimensional multitrap system where no   external fiel is present   
 and pay  special attention to      the transmission of  the diffusing 
particles through the  imperfect traps.   We discuss here the
case in which an external field is included to each trap and find not only the 
transmission  but also  the energy associated 
with the diffusing  particles in the presence and absence of such a  
field. From the energy we find the  specific
heat $C_h$  and show that for certain   values 
of the  parameters associated with the  multitrap system it 
behaves in a manner which is suggestive of   phase transition. 
Moreover, this phase transition is
demonstrated not only through the conventional single peak at  which the 
specific heat
function is undifferentiable  but also through the less frequent  phenomenon 
of double
peaks.      }

\end{abstract}

\maketitle

\bigskip \bigskip  

\section{Introduction}

The diffusion of classical particles in the presence of  traps
 have been studied by various authors 
 \cite{Abramson,Smol,Havlin,Noyes,Budde}. The idea of imperfect  traps 
 was introduced in \cite{Collins} and elaborated further by others
 \cite{Taitelbaum,Condat} which study, especially,  the nearest neighbours 
 distance in the
 presence of a single imperfect trap. We note that the imperfect trap  
 may
represent a model for many real physical situations that are not easy  to tackle
directly. Among these  one may enumerate  all the reactions, including the
chemical ones,  in which the results for the reacting particles can  not be 
predicted 
beforehand. That is, these reactions  may result in  either the elimination of
some of the reacting particles or changing the values of some physical
parameters that are associated with them. Among these parameters one may
mention, for example, the transmission and absorption  coefficients of the
diffusing particles,  their energy and  the other variables derived from 
it such
as the specific heat, the entropy etc. In \cite{Bar1,Bar2} the transmission
properties of these  particles in the presence of $N$ imperfect traps, 
without  any external field,  were intensively discussed for both cases of 
large and small $N$. It was found
\cite{Bar1} that the larger is $N$ the higher is the transmission 
 so that for $N \to \infty$  the probability that all the particles diffuse  
through all the traps tends to unity. \par
In this work we include an external field to each trap and find the effect of 
it upon the
diffusing  particles. We note that the case of a bias field in the presence
of a single imperfect trap at the origin was discussed by Condat {\it et al} 
 in \cite{Condat} which
studied among other things the effect of a uniform field on the 
distribution of nearest neighbours
distances.   
    In this work we introduce the bounded multitrap system and 
    discuss not only the trasmission that results from
including an external field to each trap but also  
 the energy associated with the diffusing particles. 
 This energy    is discussed 
 for both cases of the existence  and absence of a field and  we show that 
  it   is   highly 
 sensitive to  the presence of it so that    
     increasing its   intensity   by a
 small amount may  result in  a disproportionally large change in the 
 energies of the
 diffusing particles. \par
 We note that although the physical situation is not strictly of the equilibrium
 kind we follow  the tendency  of  many authors \cite{Haken} 
 to
 discuss phase transition and specific heat even at situations which are far
 from equilibrium.   In this context we note that it is well known that
 diffusion on a lattice can be mapped onto an equilibrium polymer problem
 \cite{Gennes}. Note that the
 imperfect traps, which introduces the diffusivity in the multitrap
 system,  are characterized here by a rather small constant
 trapping rate
 (denoted by $k$) as may be seen from the numerical work 
  in which we assume 
 $k=1$  compared to the value of $k=\infty$  for the  ideal traps. Thus, the
 bounded multitrap array deviates only slightly from   equilibrium.  
 Also, the fields added to this equations are
 entirely {\it uniform} as realized from the numerical part of this work. 
     We may, 
 therefore,  discuss the specific heat and other thermodynamical variables. \par  
    We show in the following that the mentioned  large changes in the 
    energy entail,  for certain
 values of the parameters associated with the multitrap system,  
  corresponding
 large and even discontinuous   changes   in the specific heat $C_h$ 
   derived from
 it.  That is,  
 the specific heat actually goes through phase transition which is demonstrated
 not only through the conventional undifferentiable  single peak but also 
 through
 double peaks \cite{Fendley,Bar4,Neda,Kim,Tanaka,Pagliuso,Ko}.    
  We note that   
double peak phase transitions were    found \cite{Bar4} in the 
analogous  
quantum 
one-dimensional bounded multibarrier potential.  Lieung and Neda \cite{Neda} and 
also Kim {\it et al} \cite{Kim} have found
 double peaks in the form of the responsive curves which are apparently 
 associated with
 dynamically induced phase transitions. Tanaka {\it et al} \cite{Tanaka} have found
 such double peaks   in antiferromagnetic materials corresponding to
 magnetic phases where the external magnetic field has a corresponding role 
 to the parameter $c$ here.  Ko and Asakawa \cite{Ko} have found also double peaks 
  with regards  to the phases of the quark-gluon plasma which may  be thought of
  as a large number of interactions in a bounded region.     
 \par  
In Section 2 we represent the appropriate terminology and terms associated with
the one-dimensional bounded imperfect multitrap system as in \cite{Bar1,Bar2}.
We then show, using  the $N$ $(2X2)$ transfer matrices  
method \cite{Merzbacher,Yu},     that for certain values of the 
parameters associated with 
the multitrap system the transmission probability  
through all the traps  tends to unity  in the presence of 
an external field. This was shown \cite{Bar1,Bar2},  
 for the absence of field,  by  applying  the 
same method. 
 Using the results of Section 2 we calculate in 
Section 3  the energy associated with the diffusing particles  for both
cases of the presence and absence of a bias  field. For this we use 
 the single $(4NX4N)$ transfer matrix (as done in \cite{Bar1})  for 
calculating 
the coefficients of the density of the diffusing particles. 
 In Section 4 we
calculate the relevant specific heat $C_h$  and
show that for certain values of the   variables  associated with the bounded
system 
 it   behaves in  a manner which is suggestive of  phase transition. 
This  is demonstrated in the presence of an undifferentiable 
discontinuous peaks  in the  curves  of $C_h$ as function of the temperature. Moreover, as noted, 
we find for some values of the relevant variables  
that this phase transition is demonstrated in the form of double peaks. 
    We then 
 calculate the relevant critical exponents associated with these phase
 transitions.  In Section 5 we conclude with a brief summary.

\section{The one-dimensional bounded imperfect multitrap system}
We assume that  the imperfect traps,    through which  
the particles diffuse,  
 are all characterized by  the same   width that depends 
upon their  number $N$  and
the  length $L$ of the system. 
 That is, denoting the total width of  the $N$ traps by $a$ and the total 
 interval among them  by $b$  we can see that the width of each  is
 $\frac{a}{N}$ and the interval between any two neighbours  is
 $\frac{b}{N}$. We define as in  \cite{Bar1,Bar2}  the ratio 
   $c=\frac{b}{a}$ and express $a$ and
 $b$ in terms of $c$ and the total length $L=a+b$ as 
 $ a=\frac{L}{(1+c)}, \ \ \ \ b=\frac{Lc}{(1+c)} $.  
The one-dimensional imperfect multibarrier system is assumed to be arrayed along
the positive $x$ axis so that it begins from the point 
 $x=\frac{b}{N}=\frac{Lc}{N(1+c)}$  
 and  ends  at $x=L$.  
We also assume that  at $t=0$ the particles are concentrated, except for an 
initial configuration $f$,  at the left hand faces  of the $N$ traps. 
 The locations of  
these   faces,    
 according  to  the mentioned arrangement of the multitrap system,  is given by 
  \begin{equation} \label{e1} {\grave
x_m}=m\frac{Lc}{N(1+c)}+(m-1)\frac{L}{N(1+c)}, \ \ \ \ \ \ \ 1 \le m \le N
\end{equation}  
 It is assumed  
that each trap have  the same  field associated with it so that 
any    particle that approach  it may be, depending on the
direction of the field,  either attracted or repelled by it.   
The corresponding initial and boundary balue problem associated with 
the one-dimensional bounded
imperfect $N$ trap system is \cite{Condat,Bar1,Bar2}
\begin{eqnarray} && 
\rho_t( x, {\grave x_m},t)=D\rho_{ x x}( x,{\grave x_m},t) 
-V\rho_{ x}( x, {\grave x_m},t), \ \ \ \ \ \ \  0 \le x,{\grave x_m} \le L, 
 \nonumber \\ && \ \ \ \ \ \ \ \ \ \ \ \ \ \  \ \ \ \ 1 \le  m \le N,  
  \ \ \  \ \ \ \ \  t>0  \nonumber \\ 
&& \rho(  x, {\grave x_m},0)=e^{-\frac{V {\grave x_m}}{2D}}
\delta( x- 
{\grave x_m})+f( x, {\grave x_m},0), \ \ \ \ 0 \le x,{\grave x_m} \le  L, 
\label{e2}  \\ && \ \ \ \ \ \ \ \ \ \ \ \ \ \ \  \ \ \ \     1 \le m \le N , 
                 \ \ \ \  \ \ \ \ \ t>0                          \nonumber  \\
&& \frac{\partial\rho}{\partial  x}|_{ x=  {\grave x_m}}=
(k+\frac{V}{D})\rho({\grave x_m},t),  \ \ \ \ \ \  \ \  
 0 \le {\grave x_m} \le L, 
\ \ \ \ \ \ \ \ \ \ \  1 \le m \le N, \ \ \ \   
   \ \ \ \ t>0,                              \nonumber \end{eqnarray}  
    where ${\grave x_m}$ denotes   the left hand face  of the $m$-th trap 
    (see Eq (\ref{e1})).       
        $\rho( x, {\grave x_m},t)$ is the density of the 
diffusing particles and 
$D$ is the
diffusion constant which is considered to  have two values; $D_i$ 
inside the traps and  $D_o$
 outside them  \cite{Bar1,Bar2}. We note that one may generally
find in the literature values of $D$ in the range $0.9 \ge D \ge 0.3$ (see, for
example, p. 337 in \cite{Reif} where it is noted that $0.5 \frac{cm^2}{sec}$ is
the order of magnitude of $D$ at room temperature and atmospheric pressure). We
have assigned here for $D_o$ and $D_i$ the respective values of $0.8$ and $0.4$.  
$\rho_t( x, {\grave x_m},t)$ and $\rho_{ x x}( x, {\grave 
x_m},t)$ 
are respectively the first time derivative  and the
second spatial derivative.  The parameter $V$ is the diffusion
velocity that results from the presence of the field and its magnitude and 
 sign represent respectively the intensity and direction of the field. 
The second equation of the set
(\ref{e2}) is the initial condition at $t=0$ the first term of it 
 signifies,  through the $\delta$,  that
there is an initial  concentration of the particles at 
 the left hand face of the 
trap located at  
$x={\grave x_m}, \ \ 1 \le m \le N$.  The second 
term at the right hand side  denotes  an additional 
initial configuration of the density that depends upon $ x$ and
$ {\grave x_m}$.  The third equation is the
boundary value condition  which introduces the  velocity $V$ 
that results from  the field. The
parameter $k$ denotes the degree of imperfection of the traps (in \cite{Condat}
it is termed the trapping rate) so that in the
limit $k \to \infty$ the trap becomes ideal in which case any particle that
approaches it is absorbed. We follow the procedure  in \cite{Bar1,Bar2} 
for $V=0$   and  
  decompose the set (\ref{e2})  into two secondary sets as follows
\begin{eqnarray} && 
\rho_t( x, {\grave x_m},t)=D\rho_{xx}(x,{\grave x_m},t) 
-V\rho_{ x}( x,{\grave x_m},t), \ \ \ \ \ \ \ \ \ \ 0 \le x,x_m \le L, 
 \nonumber \\ 
 && \ \ \ \ \ \ \ \ \ \  1 \le m \le N, \ \ \ \ \ \   t>0,    \nonumber \\ 
&& \rho( x,{\grave x_m},0)=e^{-\frac{V {\grave x_m}}{2D}}
\delta(x-
{\grave x_m}),   \ \ \ \  0 \le x,x_m \le L, \ \ \ \ \ \ \ \ 
 1 \le m \le N \label{e3} \\
&& \frac{\partial\rho}{\partial  x}|_{x={\grave x_m}}=
(k+\frac{V}{D})\rho({\grave x_m},t),  \ \ \ \ \ \  \ \ \  0 \le x_m \le L, 
 \ \ \ 
1 \le m \le N, \ \ \ \ \ \ \ t>0    \nonumber \end{eqnarray} 

\begin{eqnarray} && 
\rho_t( x, {\grave x_m},t)=D\rho_{ x x}( x, 
{\grave x_m},t), \ \ \ \ 
\ \ 0 <  x,{\grave x_m} \le L, \ \  \ \ \  1 \le m \le N, \ \ \  t>0 \nonumber \\ 
&& \rho( x, {\grave x_m},0)=f( x, {\grave x},0), 
  \ \ \ \ \ \ \ \ \ \ \ \ \ \ \ 
0 < x,{\grave x_m} \le L, \ \ \  1 \le m \le N \label{e4} \\
&& \rho( {\grave x_m},t)=0, 
 \ \ \ \ \ \  \ \ \  \ \ \ \ \ \ \ \ \ \ \ \ \ \ \ \ \ \ \ 
0 < {\grave x_m} \le L,  \ \ \ 1 \le m \le N, \ \ \ \ t>0  \nonumber \end{eqnarray} 

The set (\ref{e4}) is the ideal trap problem in which no external field is
present as may be realized from the absence of $k$ and $V$. 
  The set (\ref{e3}) is the imperfect trap problem  which includes the external
  field and its  initial
condition contains only the first term  of the corresponding  condition of the 
general set (\ref{e2}). The solution of 
(\ref{e2}) may be written  
 as in \cite{Bar1,Bar2} (see Eq (4) there),  
\begin{eqnarray}  && \rho( x, {\grave x_m},t)=
A\rho_1( x, {\grave x_m},t)+B\rho_2( x, {\grave x_m},t), \ \ \ \ \ \
  0 \le x,{\grave x_m} \le L,  \label{e5} \\ && \ \ \ \ \ \
  \ \ 1 \le m \le N, \ \ \ \ \ \ \ \ t>0 \nonumber   \end{eqnarray}
where $\rho_1( x, {\grave x_m},t)$ is the solution of the initial 
and boundary value problem
of the set (\ref{e3}) and $\rho_2( x, {\grave x_m}, t)$ is that of 
(\ref{e4}). The ideal trap
problem of  (\ref{e4}) is the same as that discussed  in
\cite{Bar1,Bar2}  so, following the discussion there, 
  we   write for $\rho_2( x, {\grave x_m},t)$  
\begin{equation} \label{e6} \rho_2( x, {\grave x_m},t)=
\sin(\frac{\pi  x}{ {\grave x_m}})
\exp(-\frac{Dt\pi^2}{ {\grave x_m}^2}), \ \ \    0 \le x,{\grave x_m} \le L, 
\ \ \ 
1 \le m \le N, \ \ \ \ t>0  \end{equation} 
Regarding  $\rho_1( x, {\grave x_m},t)$ we may follow Condat {\it 
et al}  in \cite{Condat},  which
discuss the problem (\ref{e3}) for the single trap at the origin,  and write 
the solution for
the multitrap case as  
\begin{eqnarray}   && \rho_1( x, {\grave x_m},t)=
\exp(\frac{V( x- {\grave x_m})}{2D}-
\frac{V^2t}{4D})(\frac{e^{-\frac{( x- {\grave x_m})^2}{4Dt}}}
{\sqrt{\pi
Dt}} - (k+\frac{V}{2D}) \cdot \nonumber \\ &&  \cdot 
\exp(Dt(k+\frac{V}{2D})^2+(k+\frac{V}{2D})( x- {\grave x_m}))
erfc(\frac{( x- {\grave x_m})}{2\sqrt{Dt}}+(k+\frac{V}{2D})\sqrt{Dt})), 
\label{e7} \\ && 
\ \ \ 0 < x,{\grave x_m} \le L, \ \ \ 
1 \le m \le N, \ \ \ \ \ \ t >0,   
\nonumber \end{eqnarray}
  where    $erfc(x)$  is the  complementary error 
function defined as   \cite{Abramow}
$  erfc(x)=\frac{2}{\sqrt{\pi}}\int_x^{\infty}e^{-u^2}du $.  
 In
order to adapt the general solution (\ref{e5})   
to the bounded $N$
multitrap system we may use, as in \cite{Bar1,Bar2},  either the $N$  
$(2X2)$  transfer matrices 
method
\cite{Merzbacher,Yu} or the equivalent 
single  $(4NX4N)$ transfer matrix formalism.  We use in
 this section the $N$ $(2X2)$ transfer matrices method for $V \ne 0$  and write 
 the general  matrix equation  \cite{Bar1,Bar2}  
 \begin{eqnarray}  &&
\left(
\begin{array}{c} A_{2N+1} \\ B_{2N+1} \end{array}
\right)={\cal T}(a+b){\cal T}(\frac{(N-1)(a+b)}{N}){\cal T}(\frac{(N-2)(a+b)}{N})\ldots  
\label{e8} \\ && \ldots 
 {\cal T}(\frac{n(a+b)}{N}){\cal T}(\frac{(n-1)(a+b)}{N}) \ldots
    {\cal T}(\frac{2(a+b)}{N}){\cal T}(\frac{a+b}{N})\left(
\begin{array}{c} A_1 \\ B_1  \end{array}
\right),  \nonumber \end{eqnarray}
where the  $A$'s and $B$'s are respectively the coefficients of the 
imperfect  and ideal 
 trap components 
 from Eq (\ref{e5}).    The  $N$  ${\cal T}$'s in Eq (\ref{e8}) are  all 
  two-dimensional transfer matrices that differ from each other by   the
  value of $x$ only \cite{Bar1}.   Thus,  a representative one which  relates 
   the two faces of the same  trap that are located, for example, 
     at $x=x_1$ and $x=x_2$, where  $x_2 > x_1$,  may be written as \cite{Bar1,Bar2} 
  \begin{equation} \label{e9} {\cal T}(x_1,x_2)=\left[
\begin{array}{cc} {\cal T}_{11}(x_1,x_2)&{\cal T}_{12}(x_1,x_2) \\ {\cal T}_{21}(x_1,x_2) &
{\cal T}_{22}(x_1,x_2)\end{array} 
\right], \end{equation}  
where \cite{Bar1,Bar2}  \begin{equation} \label{e10} {\cal T}_{11}(x_1,x_2)=\frac{\alpha(D_o,x_1,t)
\alpha(D_i,x_2,t)}{\alpha(D_i,x_1,t)
\alpha(D_o,x_2,t)} \end{equation}    \begin{equation} \label{e11} 
{\cal T}_{12}(x_1,x_2)=0 \end{equation}  
  \begin{eqnarray} && {\cal T}_{21}(x_1,x_2)=
\frac{\eta(D_i,x_2,t)}
{\eta(D_o,x_2,t)}(\frac{\xi(D_o,x_1,t)}
{\eta(D_i,x_1,t)} -\frac{\alpha(D_o,x_1,t)\xi(D_i,x_1,t)}
{\alpha(D_i,x_1,t)\eta(D_i,x_1,t)}) + \nonumber  \\ &&
+\frac{\alpha(D_o,x_1,t)}
{\alpha(D_i,x_1,t)}(\frac{\xi(D_i,x_2,t)}
{\eta(D_o,x_2,t)}- \frac{\alpha(D_i,x_2,t)\xi(D_o,x_2,t)}
{\alpha(D_o,x_2,t)\eta(D_o,x_2,t)}) \label{e12} \end{eqnarray}
\begin{equation} \label{e13}  {\cal T}_{22}(x_1,x_2)=\frac{\eta(D_o,x_1,t)
\eta(D_i,x_2,t)}{\eta(D_i,x_1,t)
\eta(D_o,x_2,t)}   \end{equation}   
The parameters  $\alpha$, $\xi$, and $\eta$  are 
given  by  (compare with the $V=0$ case in \cite{Bar1,Bar2})   
\begin{eqnarray}   && \alpha(D,x,t)=\rho_2(x,t)=
\exp(\frac{V(x-{\grave x_m})}{2D}-\frac{V^2t}{4D})(\frac{e^{-\frac{(x-{\grave 
x_m})^2}{4Dt}}}
{\sqrt{\pi
Dt}} - (k+\frac{V}{2D}) \cdot \label{e14} \\ &&  \cdot 
\exp(Dt(k+\frac{V}{2D})^2+(k+\frac{V}{2D})(x-{\grave x_m}))
erfc(\frac{(x-{\grave x_m})}{2\sqrt{Dt}}+(k+\frac{V}{2D})\sqrt{Dt}))
\nonumber \end{eqnarray}
\begin{eqnarray}  && \xi(D,x,t)= 
\frac{\partial \alpha(x,t)}{\partial x}=
\exp(\frac{V(x-{\grave x_m})}{2D}-\frac{V^2t}{4D})(\frac{e^{-\frac{(x-{\grave 
x_m})^2}{4Dt}}}
{\sqrt{\pi
Dt}}((k+\frac{V}{2D})-  \nonumber \\ && -  \frac{(x-{\grave x_m})}{2Dt})  
-(k+\frac{V}{2D})^2 \cdot 
\exp(Dt(k+\frac{V}{2D})^2+(k+\frac{V}{2D})(x-{\grave x_m})) \cdot \label{e15} \\ && 
\cdot erfc(\frac{(x-{\grave x_m})}{2\sqrt{Dt}}+ (k+\frac{V}{2D})\sqrt{Dt}))
\nonumber \end{eqnarray}
\begin{equation} \label{e16} \eta(D,{\grave x_m},t) =-\frac{\pi}{{\grave x_m}}
e^{-(\frac{\pi}{{\grave x_m}})^2Dt} \end{equation}

In order to find the density $\rho(x,\grave x_m,t)$ from Eq (\ref{e5}) 
at each point $x$ in the multitrap system we have to
determine the coefficients $A$ and $B$ of the imperfect  and ideal trap
components at this point. If, for example, this point happens  to fall at  the
$m$-th  trap  then 
 one have to multiply $m$ transfer matrices each of the kind given by Eqs
(\ref{e10})-(\ref{e13}).  We denote the  components of the
two-dimensional matrix that results from such a  product 
  by  $ T_{m_{11}}$, $ T_{m_{12}}$, 
$  T_{m_{21}}$
and $ T_{m_{22}}$ and those of any one of the multiplied  $m$ 
  matrices  by 
${\cal T}_{11}(l)$, ${\cal T}_{12}(l)$, ${\cal T}_{21}(l)$
and ${\cal T}_{22}(l)$ where $1 \le l \le m$.  
  Thus,  as realized  from  Eqs  
(\ref{e10})-(\ref{e13})  (and from 
\cite{Bar2}  for $V=0$),    
 the components $ T_{N_{11}}$, 
$ T_{N_{12}}$, $ T_{N_{21}}$  and $T_{N_{22}}$  may 
be expressed recursively as 
\begin{eqnarray} &&  T_{N_{11}}= T_{(N-1)_{11}}{\cal T}_{11}(N) =\ldots=
{\cal T}_{11}(N){\cal T}_{11}(N-1) \ldots
{\cal T}_{11}(2){\cal T}_{11}(1) \nonumber \\ &&
 T_{N_{12}}= T_{(N-1)_{12}}=\ldots= T_{2_{12}}={\cal T}_{12}(1)=0 
\nonumber \\ 
&&  T_{N_{21}}= T_{(N-1)_{21}}{\cal T}_{22}(N)+
 T_{(N-1)_{11}}{\cal T}_{21}(N) \label{e17} \\ 
&&  T_{N_{22}}= T_{(N-1)_{22}}{\cal T}_{22}(N) =\ldots=
 {\cal T}_{22}(N){\cal T}_{22}(N-1)\ldots 
{\cal T}_{22}(2){\cal T}_{22}(1), \nonumber \end{eqnarray}

It is found that the component $ T_{N_{11}}$ tends to unity for growing
values of the variables $N$ or (and) $c$. That is, the larger is either the
number of traps or the intervals among them (or both)  the higher is the tendency 
 of the component $ T_{N_{11}}$ to unity. Note
that the same result have been found also in \cite{Bar1} for $V=0$ 
 with respect to the same variables $N$ and $c$ (see the discussion there
after Eq (30)). We also find regarding the variable $V$   that the
lower its value is  the smaller  
  becomes $ T_{N_{11}}$.  
 The same result of  small $ T_{N_{11}}$   
 is obtained also for 
  negative  $V$.   We  also find  
   that $ T_{N_{11}} \to 0$ for  growing    values of the total length 
   $L$ and tends to unity for small values of it. 
 Also,  $ T_{N_{11}}$ increases  for larger values of
either the time $t$ or  ${\grave x_m}$.   \par
Regarding the component   $ T_{N_{22}}$  we find that it does 
not assume values 
outside the range  $(0, 1)$.   Also,   its factors ${\cal T}_{22}$ satisfy
$\lim_{N \to \infty} {\cal T}_{22} =0$ for $x \approx 0$ and 
$\lim_{N \to \infty} {\cal T}_{22} =1$ for $x \approx L$   
  (see   in \cite{Bar1}  Eq (26)   and the unnumbered equation after Eq (27)). 
  Thus, at large   $N$,    the product 
  $ T_{N_{22}}$ also tends to either unity or 0. 
   We 
  also find for small $c$   
that the larger is either the number of traps $N$ or the time $t$    
 the more apparently  $T_{N_{22}}$  
tends  to zero and this holds  also for small values of 
 $L$.  When, however, $c$    increases  $T_{N_{22}}$ clearly tends to unity. 
 Note that $ T_{N_{22}}$ does not depend upon the
variable $V$ as seen from Eqs (\ref{e13}) and (\ref{e16}). \par
 The component $ T_{N_{21}}$  may generally have any  value from  $(-\infty,
 +\infty)$  but there are specific ranges
of $N$, $L$, $c$, $V$, $k$ and $t$ for which $ T_{N_{21}}$ tends to zero. 
Thus, it is found  for growing  $c$  that  $ T_{N_{21}}$   decreases 
   fastly to    zero if the  increasing values of $c$ are small 
  and  slowly  if these  values are large. For example,  
increasing $c$ in
the range $20 \le c \le \infty$   
causes $ T_{N_{21}}$ to decrease so slowly that it may be regarded as almost 
constant over this range. 
We find that $ T_{N_{21}}$  assumes very high  values for either 
large $N$ or  small $L$ but when $L$
 grows to a value which is comparable to that of $N$ the component  
$ T_{N_{21}}$ may even decrease to zero. It is also found that the  larger
is $V$ the smaller become the corresponding values of $ T_{N_{21}}$. For example,
for $V \ge 24$ $ T_{N_{21}}$ assumes very small values of the order of $10^{-11}$
but when  $V$ decreases below   20 the value  of $ T_{N_{21}}$ 
 becomes  very large. Negative
values of $V$ have similar  effect on  $ T_{N_{21}}$ as the corresponding
positive ones.  Also,  $ T_{N_{21}}$  increases for larger values of 
the time $t$ and
decreases for small    ${\grave x_m}$. 
\par 
We, thus,   find   that  there are 
ranges of 
$N$, $L$, $c$, $V$, $k$ and $t$  for  which the components of the total matrix 
$T_N$,  that result from the $N$ products at the right hand side of 
Eq (\ref{e8}), 
assume the values  of $ T_{N_{11}} = 1$,  $ T_{N_{21}} = 0 $ and 
$ T_{N_{22}} = 1$.  In this case   all the diffusing particles pass
through all the traps in which case the transmission is maximal. That is,  
the ideal and imperfect trap components of the initial density  do not change
by the presence of either the traps or the fields (or both) 
 in which case the product of the $N$
transfer matrices at the right hand side of Eq (\ref{e8}) results in the
two-dimensional unity matrix.  
  This is seen in Figure 1 which is composed of the three
Panels $A-C$ which respectively show three-dimensional surfaces of 
 the components $ T_{N_{11}}$,
$ T_{N_{21}}$ and $T_{N_{22}}$ as functions of $c$ and $N$. All the three
Panels are drawn for $V=15$, $k=1$, $D_o=0.8$,   $D_i=0.4$, $t=1$ and $L=5$ and for
the same ranges of $15 \ge N \ge 1$ and  $300 \ge c \ge 20$. Panel $A$ shows 
the component $ T_{N_{11}}$ and one may see that for large values of $c$ and $N$ 
 $ T_{N_{11}} \to 1$. The same result is obtained also in Panel $C$ for the
 component  $ T_{N_{22}}$ whereas in Panel $B$ we see that for large values
 of $c$ and $N$ the component  $ T_{N_{21}}$ tends to zero. 
Thus, subtituting these values in Eq (\ref{e8}) and using the fact 
 that  one always have \cite{Bar1,Bar2} $ T_{N_{12}} =0$   we obtain  
\begin{equation}  \label{e18}
\left(
\begin{array}{c} A_{31} \\ B_{31} \end{array}
\right)=   \left(
\begin{array}{cc}  1 & 0 \\ 0 & 1   \end{array}
\right) \left(
\begin{array}{c} A_1 \\ B_1  \end{array}
\right),   \end{equation}
 That is, all the particles that approach the
multitrap system pass through it without any decrease in either the ideal or
the imperfect trap components of the density. Note that for $V=0$ we have shown 
in \cite{Bar1}  this unity value of the  transmission for the imperfect trap 
component of the 
density  (see Eq (28) there) and   the same result is obtained in 
\cite{Bar2}
for the ideal trap component.

\section{The energy associated with the bounded multitrap system}

We, now,  discuss  the energy associated with the diffusing particles in the
presence of an external field. 
 We do this by   following   the conventional discussion 
one may find in the literature regarding diffusive systems in the absence of
external fields (see, for example, \cite{Reif}). The presence of the field
introduces an additional source of energy (besides that related to the diffusion
through the  traps) which
must be taken care of by  adding an extra  term (to the kinetic energy)  
that depends upon the velocity $V$. 
We, thus, assume that the total 
energy is composed of two parts; kinetic and potential where the former results
from the diffusive motion  and the external field and the later from the 
presence of the traps. In the
absence of any external field the  particles diffuse with an average 
 diffusion
velocity ${\bar v_D}$ given, for the one-dimensional case,  by \cite{Reif} 
${\bar v_D}=\sqrt{\frac{2D}{t}}$ where $D$ is the diffusion
constant.  Note that since  the two densities  inside and outside the traps 
   satisfy $D_i \ne D_o$ then also  the  diffusion velocity ${\bar v_D}$ and 
   the general density $\rho$ from Eq (\ref{e5}) satisfy 
   ${\bar v_{D_i}} \ne {\bar
   v_{D_o}}$ and $\rho_{D_i} \ne \rho_{D_o}$.   
  As remarked, the natural diffusive motion  for $V=0$   is
 towards the positive  $x$ axis so  when a field is 
 present and points  in that direction it  accelerates the motion of
the particles    or decelerates it  if
it is   oppositely  directed.  Thus,  we may write the kinetic
energy of the  diffusing particles as   \begin{eqnarray}  
&& E_{K_{V \ne 0}}(x,{\grave x_m},t)=\frac{1}{2}\rho(x,{\grave x_m},t)(v_D^2 \pm V^2)=
(A\rho_1(x,{\grave x_m},t)+B\rho_2(x,{\grave x_m},t)) \cdot \nonumber 
\\ && \cdot 
(\frac{D}{t} \pm \frac{V^2}{2}), \ \ \ \ \ \ \  
 0 \le x,{\grave x_m} \le L, \ \ \ \ 1 \le m \le N, \ \ \ \ t>0,  \label{e19} 
\end{eqnarray}
where  $\rho_2(x,{\grave x_m},t)$ and $\rho_1(x,{\grave x_m},t)$ are given by
Eqs (\ref{e6})-(\ref{e7}). 
$V$ is, as remarked, the velocity that results from the external field and the
plus and minus signs in front of $\frac{V^2}{2}$ denote respectively that the
 kinetic energy due to the field is either added  for $V>0$ or subtracted when $V<0$ 
 from that 
 due to ${\bar v_D}$. 
 Note that for large $t$ the kinetic energy from the last equation 
 becomes zero  since in this case both $\rho_1$ and $\rho_2$ vanishes
as realized from Eqs (\ref{e6})-(\ref{e7}).  \par
By following  the conventional discussion of the energy in classical
diffusive systems \cite{Reif}  we may conclude that   the
force that acts on the particles is  related to the potential energy which
originates from the presence of the traps. This force is assumed to 
 be proportional to
the trapping rate $k$ so for very large $k$ (ideal traps) it assumes  maximal
values and  for $k=0$  (absence of traps) it vanishes.
Also, since we have always assigned throughout this work  a rather small value 
of unity for $k$ which means that the imperfect traps have  weak 
influence upon the particles we correspondingly assume an inverse
proportionality of the force to the {\bf squared} distance of the particles 
from the traps. This means, as remarked,  that the particles feel the effect of
the traps only at small distances from the traps. 
 Thus,  one
may write the force on any particle that results from the trap as
$F(x,{\grave x_m})=-\frac{gk}{(x-{\grave x_m})^2}$, where    $g$  
 is the proportionality constant and the
minus sign indicates  an attractive force.  
Thus,  assuming that each
trap serves as a central force source one may find the potential energy from 
\begin{eqnarray}  && E_P(x,{\grave x_m})=-\int_{(x_r-{\grave x_m})}^{(x-{\grave x_m})}
F(x,{\grave x_m})dx=
-\int_{(x_r-{\grave x_m})}^{(x-{\grave x_m})}
(-\frac{gk}{(x-{\grave x_m})^2})dx=  
\nonumber  \\ && =
-gk(\frac{1}{(x-{\grave x_m})}-\frac{1}{(x_r-{\grave x_m})})=   
-gk(\frac{1}{(x-{\grave x_m})}), \label{e20} \\ 
&& \ \ \ \ \ \ \ \ \ 0 \le x,{\grave x_m} \le L, \ \ \ \ 1 \le m \le N,  
\nonumber  \end{eqnarray}
where we assume that the reference point $x_r$ is  at infinity.   Thus,
 we may write the total energy of the diffusing
particles in the presence of  field as \begin{eqnarray} && 
E_{total_{(V \ne 0)}}(x,{\grave x_m},t)=
E_{K_{(V \ne 0)}}(x,{\grave x_m},t)+E_P(x,{\grave x_m}) \label{e21} \\
 && 0 \le x,{\grave x_m} \le L, \ \
\ \ 1 \le m \le N, \ \ \ \ t>0,   \nonumber \end{eqnarray} 
where $E_{K_{(V \ne 0)}}(x,{\grave x_m},t)$ and $E_P(x,{\grave x_m})$ are given
 respectively by Eqs (\ref{e19}) and (\ref{e20}). 
In the absence of an external field the total energy is 
\begin{eqnarray} && E_{total_{(V=0)}}(x,{\grave x_m},t)=
E_{K_{(V=0)}}(x,{\grave x_m},t)+E_P(x,{\grave x_m})=
\frac{1}{2}\rho_{V=0}(x,{\grave x_m},t)\bar v_D^2+  \nonumber  \\ && + 
E_P(x,{\grave x_m})=\frac{1}{2}(A\rho_{1_{V=0}}(x,t)+B\rho_2(x,{\grave
x_m},t))\cdot \frac{D}{t}, \label{e22} \\ 
&&   0 \le x,{\grave x_m} \le L, \ \ \ \ 
 1 \le m \le N, \ \ \ \ t>0,   \nonumber 
\end{eqnarray} where $\rho_{1_{V=0}}$ is the imperfect trap component of the 
density for $V=0$ and is given as 
$\rho_{1_{V=0}}=erf(\frac{x}{2\sqrt{Dt}})+exp(k^2Dt+kx)\cdot
erfc(k\sqrt{Dt}+\frac{x}{2\sqrt{Dt}})$ (see Eq (6) in  \cite{Bar1}). 
 $\rho_2(x,{\grave x_m},t))$ and $ E_P(x,{\grave x_m})$ are given respectively by
 Eqs (\ref{e6}) and (\ref{e20}).   The coefficients $A$ and $B$ in the last two
equations are numerically determined in this section 
 from the single $(4NX4N)$ matrix method (see the discussion after Eq (29) in 
  \cite{Bar1}).\par
We now show  that increasing $|V|$  by even a small amount may change 
 the energy in such an unexpected manner 
     that it results,   as seen in the
following section, in a phase transition of the corresponding 
specific heat $C_h$. 
This may realized form Panels $A$-$D$ of Figure 2 which all show  three
dimensional surfaces of the energy  $E$ as function of $x$ and  $c$. 
 Panels $A$ and $B$  
are both drawn for $N=2$,  ${\grave x_m}=\frac{Lc}{N(1+c)}$,  $k=t=g=1$, $L=30$,  $D_o=0.8$, $D_i=0.4$,  
  $20 \ge c \ge 0.5$,  $40 \ge x \ge 0$ and differ by the value of $V$ 
   which is 2 for Panel $A$ and 5 for  $B$. 
Thus, by comparing between them  one may realize that increasing $V$ by 
only 3 units
causes to a disproportionally large increase  of $E$  
  from $|E| \approx 12$ in Panel $A$ to $E \approx 14000$ in $B$. 
  This large jump of the 
energy entails   a corresponding
discontinuous change in the values of the specific heat $C_h$ which implies, as
will be shown, that it goes through a phase transition. 
The same result is obtained  also for negative $V$ but compared to 
 $V>0$  the   changes obtained are larger  and  found   at smaller  values 
 of negative $V$.    This may be  seen in Panels $C$ and $D$ of Figure 2    
 which  are both drawn     for $N=5$,    
 ${\grave x_m}=\frac{2Lc}{N(1+c)}+\frac{L}{N(1+c)}$, $k=g=t=1$, $L=30$, $D_o=0.8$, $D_i=0.4$,    
  $20 \ge c \ge 0.5$,   $40 \ge x \ge 0$  and differ by $V$ which is 
  $-0.5$ for Panel $C$ and   $-0.8$  for $D$.   
  Thus,   by comparing these two Panels one may realize 
  that decreasing $V$ by only three tenths from $-0.5$ to  $-0.8$ results 
  in a giant  change 
   of the energy from $|E| \approx 10$ in Panel $C$ to $E \approx
  10^{8}$ in $D$.  This  entails, as will be shown in the following section, a 
  corresponding phase transition of the specific heat $C_h$. One may explain
  these large changes of the energy  by
  reasoning that increasing the intensity of the field $|V|$ beyond some limit
  causes the particles to overcome any resistance 
  related to the diffusion in the presence  of traps.  Thus, 
   their energy increases 
  disproportionally to the change of $|V|$ that causes it. A similar behaviour is
  encountered in Laser tubes \cite{Haken1} when the pumping energy (field) 
  attains a limit value  which causes 
  the intensity of the produced light to increase  in a phase
  transitsional manner. \par It is expected, regarding  the dependence of the 
energy 
upon the trapping rate  $k$, that the larger  $k$ becomes  the more controlled 
will be 
 the diffusing particles by the traps in which case the kinetic (and the total) 
 energy  of these particles decrease.  This is 
  shown from recalling that for   
  $N=2, \ \ 
 V=5, \ \ k=1, \ \  t=g=1, \ \ L=30, \ \ D_o=0.8$, 
   $D_i=0.4, \ \ {\grave
 x}=\frac{Lc}{N(1+c)}, \ \ 
   20 \ge c \ge 
 0.5$ and $40 \ge x \ge 0$ we have obtained  that the larger  
 values of the 
 energy  are  
 $E \approx 14000$. Now, 
   it have been, numerically,   found (not shown) that  if $k$ 
 is raised  from $k=1$ to $k=5$,
 keeping the values 
   of all the other parameters as before, 
     the larger values of the 
 energy  are decreased to
  $E \approx 8000$. When $k=10$ we find that 
   $E \approx
 5000$ and 
  for $k=15$ the larger  $E$ further
   decrease to $E \approx 3000$. 
 A similar behaviour of decreasing
 energies for larger $k$ is found 
  also for negative $V$.  
   \pagestyle{myheadings}
\markright{SINGLE AND DOUBLE-PEAK PHASE TRANSITION.....} 
\section{Single and double-peak phase transition in the specific heat of the
bounded multitrap system} 
The average
energy from which one may derive most of the statistical mechanics variables
such as the specific heat $C_h$, the free energy $F$, the entropy $S$ etc may be
written as  
$<\! E_{total}\!>=\frac{\sum E_{total}e^{-\beta E_{total}}}{\sum e^{-\beta
E_{total}}}, $  where $\beta=\frac{1}{k_bT}$, $k_b$ is the 
Boltzman constant and
$T$ is the temperature in Kelvin units. Substituting in the former equation 
the
appropriate expression for
$E_{total}$ from Eq (\ref{e21}) or (\ref{e23}) yields respectively the average
energy for the presence or absence of an external field. 
 From the expression of the average energy  $<\! E_{total}\!>$  we obtain 
the specific heat $C_h$  \begin{equation} \label{e23} 
C_h=\frac{\partial <E_{total}>}{\partial T}=\frac{\partial}{\partial T}
(\frac{\sum E_{total}e^{-\beta E_{total}}}{\sum e^{-\beta
E_{total}}})=\frac{1}{T^2}(<E_{total}^2>-<E_{total}>^2) \end{equation} 
Figure 3 shows a three dimensional surface of  $C_h$ 
from Eq (\ref{e23}) as function of the ratio $c$ and  the temperature $T$ 
where no external
field is present  in which case the appropriate energy to be substituted in the
last equation is that from Eq (\ref{e22}). Figure 3 is drawn for 
$N=2$, $k=1$, $L=30$, $t=1$, $g=1$, 
${\grave x_m}=\frac{Lc}{N(1+c)}$,   
$20 \ge c \ge 0.5$ and $10 \ge T \ge 0.1$. 
   As seen,  the height of the surface for $C_h$ increases with
growing values of the ratio $c$ untill some maximum (not shown in the
figure).  One may also realize that at 
 small values of $c$ and $T$  the surface of $C_h$ 
jumps upward 
 to its local maximal values from which it descends in a 
similar manner to zero. 
These local maxima  are seen to be arrayed along horizontal lines which form, 
in relation
to their neighbouring lines,   a
  sharp  edge 
 which  becomes widened and  flattened as  $c$ grows. \par 
The epecific heat
$C_h$ is
certainly undifferentiable at the sharp  edge  so 
 it  goes through phase
transition at these points  \cite{Fendley,Reichl}. We calculate at the following 
the critical exponents $\chi$ \cite{Reichl} 
associated with this  and other discontinuities  of $C_h$. 
We find  that increasing  $N$  or $k$ or $g$, while keeping the 
values of the
other parameters constant,  does not cause to any
change in the form of  $C_h$ shown in Figure 3  
except to its translation from its position along the $c$ axis to a one that
tends to be aligned   along the 
$T$ axis. That
is, the same surface of $C_h$  is rotated in the $c-T$ plane for growing values 
of  $N$ or  
$k$ or $g$. If, on the other hand, the value of the total length $L$ is
simultaneously increased with that of $N$ then the remarked rotation of the
surface of $C_h$,  obtained for large values of $N$,  is avoided and this
surface remain in its form and place.      
Unlike  the  case of $V \ne 0$,   to be
discussed in the following, we find for $V=0$  that except for the 
remarked points of
discontinuity associated with  smaller values of
$c$ and $T$ there are no other  points at which $C_h$ becomes discontinuous. 
  \par 
We discuss now the specific heat $C_h$ obtained when an external field is 
included
with each trap.  It is found, as for the $V=0$  case, that the
corresponding curves  of the specific heat $C_h$ jumps abruptly from zero for
small  $c$ and $T$ to their maximal values  from which they similarly 
descend to
zero. Also, as for Figure 3, these maximal values are arranged along  
lines which form  a sharp  edge for  small  $c$ 
and $T$ which 
become widened and flattened as $c$ increases.  The specific heat 
function is  clearly 
undifferentiable  along the sharp edge  which implies, as for the 
  $V=0$  case,  that  it goes through 
phase transition 
\cite{Fendley,Reichl} at these points. But, in contrast to the former case, 
there exist 
 other points, not at small values of $c$, for  which the specific 
heat goes through phase transition. This is demonstrated  in   Figure 4 
which is drawn for $V=5$, 
 $N=2$, $k=g=t=1$, $L=30$,  ${\grave x_m}=\frac{Lc}{N(1+c)}$ and  
 which  is related  not only to the more obvious  discontinuity of
 the two spiky columns  at  small  $c$ and $T$ but also to that 
 of  the apparently continuous  surface at large $c$'s. This is
 clearly shown in Figure 5  in which we isolate from the surface of Figure 4 
  four curves of the specific heat $C_h$ as function of
 the temerature $T$ for  $c=3, 3.1, 3.2, 3.3$.    
  One can  see that each curve of the four shown 
   assumes 
 the form of  two inverted and  indented  tooth which are clearly  
    undifferentiable  and so they  constitute a 
    double-peak phase transition.  We
 thus see, as remarked, that the unexpected large change in the values of 
 the energy in
 Panel $B$ of Figure 2, which is drawn for the same values of 
 $V$, $N$, $k$, $L$, $g$, $t$,  and ${\grave x_m}$ as those of Figures 4-5, 
 is affected through the double peak phase transition of the corresponding  
 specific heat of Figure 5.   
   This depends, as noted,  upon the value of  
  ${\grave x_m}=\frac{Lc}{N(1+c)}$, so
  we expect that changing its value may result in finding double peaks phase
 transitions at other values of $c$ and $N$. This is indeed the case as  
  we find (not shown here), for example,  
 for  $N=4$,  $10.6 \ge c \ge 10$ and 
 ${\grave x_m}=\frac{2Lc}{N(1+c)}+\frac{L}{N(1+c)}$, 
 which
 is the location of the left hand face of the second trap.       \par
 
  From the former discussion  we see that for positive values of $V$ there is 
 associated 
 a single peak for the smaller values of $c$ and a double peak for some higher
 values of it. When we consider, however, negative values of $V$ we find that
 the double peaks generally emerge    for the smaller values of $c$. 
 This is
 demonstrated in the right hand Panel  of Figure 6 which shows 2 curves of the specific 
 heat $C_h$ as function of
 the temperature $T$ for $V=-1.92$, $g=k=t=1$, $L=30$, $D_o=0.8$, $D_i=0.4$, 
 ${\grave x_m}=\frac{Lc}{N(1+c)}$  and for the two 
 values of $c=0.39, 0.4$.     In this case the first peaks of the two curves 
 are 
  small compared  to the
 seconds.  
 The left hand  Panel shows 8  double 
 peaks curves of the specific heat as
 function of $T$ for ${\grave x_m}=\frac{2Lc}{N(1+c)}+\frac{L}{N(1+c)}$,  
 $N=5$,   $V=-0.37$, $g=k=t=1$ and  
 $c=1.5+0.065\cdot n, \ \ \ n=1, 2, \ldots 7$.   
  These phase transitions of $C_h$
 correspond  to the unexpected  large change of the energy which is 
 shown in Panels
 $C$-$D$ of Figure 2  for
 exactly these values of $c, N, t, g, k, {\grave x_m}$  and in the neighbourhoud
 of $V=-0.5$.   These Panels demonstrate, as remarked,  that
 slightly changing the value of $V$ in the neighbourhoud of $V=-0.5$ by only 
 three tenths changes the larger values of the energy from $|E| \approx 10^{1}$ to
 $E \approx 10^{8}$. This  change in the energy  is demonstrated in the
 double peak phase transition shown in the left hand Panel  of Figure 6.        
   Note that all the eight first peaks, as well as all the  second  peaks, 
    touch each other and seem 
 as one curve.  The appropriate energy to be associated
 with negative $V$ is that of the expression (\ref{e21}) in which one should
 take, as remarked,  the minus sign
 in front of $\frac{V^2}{2}$.   \par
  We may suggest an explanation for the occurence of the mentioned large  
 changes in the energy which entail the corresponding discontinuous peaks in the
 specific heat $C_h$. We confine our attention to the discussed examples of
 $N=2$ and $N=5$ when one respectively changes from  $V=2$  to $V=5$ and
 from $V=-0.5$  to $V=-0.8$. As remarked, the change of $V$ for $N=2$ entails a
 change in the larger values of the energy from $E \approx 12$ to 
 $E \approx 14000$. Looking at the
 expression  (\ref{e19}) for $E$ one may realize that the large increase in
 $E$ results from a corresponding increase of the imperfect density $\rho_1$
 from Eq (\ref{e7}) (the ideal  density $\rho_2$ does not depend upon $V$ and so
 it does not change with $V$ (see Eq (\ref{e6}))). As seen from Eq (\ref{e7})
 the dependence of $\rho_1$ upon $V$ is mainly exponential. 
  Thus,  when  $V$ changes from 2 to 5  we find for the ratio 
$\rho_{1_{V=5}}/\rho_{1_{V=2}}$ the value of 
$\rho_{1_{V=5}}/\rho_{1_{V=2}}=2.3096 \cdot 10^{12}$ where  
 we use the same 
values
used for all the other   parameters that  lead to Panels $A$  
and  $B$ of
Figure 2.   That 
  is, the density for $V=5$ 
has enormously grows in relation 
 to that for $V=2$.  This is to be compared, for example, to water when one 
lower its temperature  from the gaseous state to the liquid one in which case
the density of the water molecules grows in a phase transitional manner. This
behaviour is repeated when one continues to decrease the temperature to $0^0C$
from the liquid state to the solid one in which case the density of the water
molecules increases again in a phase transition manner. We have mentioned in the
previous section the example of laser tubes for which the intensity of light
increases greatly when the pumping energy (corresponding 
to $V$ here) attains a specific value.  This occurs because a macroscopic
aggregate
of atoms have been transferred by the pumping energy into the appropriate laser
state \cite{Haken}. A similar grow of the
density occurs, as remarked,  also here when one increases for $N=2$ the 
velocity $V$  
from 2 to 5. \par
The double peaks shown in Figure 6 which are associated with small negative $V$  
may be explained by noting that the external diffusion constant employed here is
$D_o$ is 0.8 (the internal diffusion constant is even lower $D_i=0.4$). That is,
when one turns on an external field which is directed opposite to the diffusive
motion then when the value of this field becomes $V=-0.8$ it actually 
neutralizes and cancels the influence of the traps on the particles so that
their energy becomes very large. \par 
The appearance of the double peaks for these values of $V$ demonstrates further
(more than the single peak) the large change that the density and the energy
have passed through  when $V$ changes as described.  We must, however,  note
that these peaks depend not only upon $V$ but also upon the other parameters,
such as $N$, $x_m$, $t$ etc, that control the behaviour of $E$. \par
 one may calculate the related critical exponents $\chi$ \cite{Reichl} 
 associated with these phase transitions by using the following equation in the
 neighbourhood of the critical temperature $T_c$  \cite{Reichl} 
 \begin{equation} \label{e24} 
 C_h(\epsilon)=A+B\epsilon^{\chi},  \end{equation} 
 where $\epsilon=\frac{T-T_c}{T_c}$ and $A$, $B$ are constant. The first order
 derivative of $ C_h(\epsilon)$  diverges at $T=T_c$ so the
 critical exponent $\chi$ may be obtained from \cite{Reichl}
\begin{equation} \label{e25}  \chi=1+\lim_{\epsilon \to 0}\frac{ln|\grave
C_h(\epsilon)|}{ln(\epsilon)}= 1+\lim_{\epsilon \to
0}\frac{ln|\frac{B}{\epsilon^{\chi}}|}{ln(\epsilon)},  \end{equation} 
where the unity term denotes the first order  derivative 
$\grave C_h(\epsilon)$ of the specific
heat with respect to $\epsilon$.  The value of $\chi$  may be obtained by 
plotting the
curve of $C_h(\epsilon)$  in the close neighbourhood of $T=T_c$ and one can 
 see from  Figures 3-6  that $\chi$   assumes
different values. Thus, assigning to the constants $A$, $B$ the respective
values of 0 and 1 and plotting, as remarked, the graphs of $C_h(\epsilon)$ in
the immediate neighbourhoud of the single peaks in Figure 3,  
which are  located at
 small  $c$ and $T$,  one may calculate 
$\chi$,   using Eqs (\ref{e24})-(\ref{e25}),  
 as $\chi \approx \frac{1}{2}$. This value changes  
with respect to the
single  peaks of Figure  4 which are also located at small  
$c$ and $T$. That is, repeating the same procedure  one may obtain  the value of $\chi \approx
\frac{3}{5}$.  For the eight  first and second  peaks  in the left hand 
Panel  of 
Figure 6  
  we find the respective 
  values of $\chi \approx \frac{1}{3}$ and  $\chi \approx \frac{1}{2}$.  
 The  $\chi$   of the 
double peaks of Figure 5 and also  of the first peaks in the right hand 
Panel  
 of Figure 6  
is $\chi \approx \frac{1}{9}$. \par
Comparing the phase transition behaviour of the bounded 
one-dimensional multitrap system to that of the corresponding quantum array of
the bounded one-dimensional multibarrier potential \cite{Bar4} one may notice
the following similarities and differences;  The specific heats of both systems 
exhibit the same discontinuous jump at small values of $T$ and $c$ but whereas in the
multitrap system $C_h$ decreases, for growing values of $T$, 
  to zero the
corresponding quantum $C_h$ does not vanish but tends, for large $T$, 
  to a finite value (see Figures 1-7 in \cite{Bar4}).   Also, the 
  phenomenon of the double peaks
  at which the specific heat $C_h$ is undifferentiable are discernable in both
  systems.   \par
 The variation of the critical 
exponent $\chi$ may be explained by  noticing from
Figures 3-6  
 that the different peaks shown in these figures  
correspond to different values of $N$ and 
 $c$.  Remembering that $N$ and $c$ 
 respectively denote the number of traps and the ratio 
  of their total
interval 
 to their total width one may realize that they, 
 actually,
  control 
   the shape and form of the multitrap system. That is, 
  Figures 
  3-6 with the different 
   values of $N$ and $c$ 
actually correspond
to  {\bf different  systems}  through which the particles pass and not to
different parameters of the same 
 system. Thus, one may expect 
  different values of 
 the critical exponent to be associated 
  with these different
systems. 
We note, however, 
  that the difference between these  
 values  is not
large. \par
  Using the  expressions (\ref{e21})-(\ref{e22}) for the energy one may obtain 
the other variables of statistical mechanics. For example,  
the free energy $F$ is calculated,  for either the presence or absence of the
external field,  from \cite{Reif} 
$F=-k_bTln(\sum e^{-\beta E_{total}})$. Using the last equation one may write 
 the entropy $S$ for the multitrap system  \cite{Reif} \begin{equation}
\label{e26}
S=-\frac{\partial F}{\partial T}=\frac{\partial}{\partial T}(k_bT
\ln(\sum e^{-\beta E_{total}}))=k_b \ln(\sum e^{-\beta E_{total}})+k_b\beta 
\frac{\sum E_{total}e^{-\beta E_{total}}}{\sum e^{-\beta
E_{total}}} \end{equation}     
 As for the specific heat $C_h$ one may draw  $S$  for different 
 values of the parameters $N$, $k$, $g$, ${\grave x_m}$,  $t$, $L$ and $c$. 
 If, for example,
 we draw  the surface of $S$   
for exactly the same values of the mentioned parameters  as those of Figure
4 one may see (not shown here)  two separate  lobes for small $c$ and $T$ which 
correspond to the two
spiky columns of Figure 4.

\section{Concluding Remarks}

We have discussed the diffusion limited problem related to the  bounded
  one-dimensional imperfect multitrap system in the presence of 
   external field $V$.   The
analytical methods previously  used \cite{Bar1,Bar2} to discuss the transmission of
the particles through this   system in the absence of an external field
were also used here in the presence of it. Thus, the $N$  $(2X2)$ transfer 
matrices 
were used,  as in \cite{Bar1},   for discussing the transmission 
through the multitrap system for $V \ne 0$ 
 and the single $(4NX4N)$ transfer matrix   for 
studying the energy and the corresponding  specific
heat $C_h$. It has been found, as for the $V=0$  case in \cite{Bar1}, 
that for certain values of the parameters associated with the  system
the transmission coefficient \cite{Bar1,Bar2} of the diffusing particles 
 tends to unity when $N$ and  $c$  become  large in which case all the 
particles
diffuse  through all the traps. This has been shown  not only for  
 positive $V$    which  pushes the   particles  
   towards  the traps but also for negative values of it that repell  
the  particles towards the negative direction of the $x$ axis. \par
The unique characteristics of the mutitrap system 
become more unexpected  regarding   the energy of the diffusing 
particles in the presence of an  external
field. Thus, it has been found that increasing  $V$, for either
positive or negative values of it,  by even a small
amount results in a disproportionally  large increase in the energy  of the
diffusing particles so that  trying to calculate the related specific
heat $C_h$ we  find that it goes through phase transition. Moreover, for certain
values of the parameters associated with the multitrap system,  such as its total
length $L$, the number of traps $N$, the ratio $c$,  the time
$t$, the location  ${\grave x_m}$ at which the particles are 
initially concentrated 
and the field $V$,    one may find  that the mentioned phase
transition is demonstrated in the form of a double peak. The value of the
related critical exponents associated with these phase transitions were found to
vary between $\frac{1}{9}$ and $\frac{3}{5}$. 
 \bigskip \bibliographystyle{plain}

\begin{figure}[hb]
\centerline{
\epsfxsize=5.5in
\epsffile{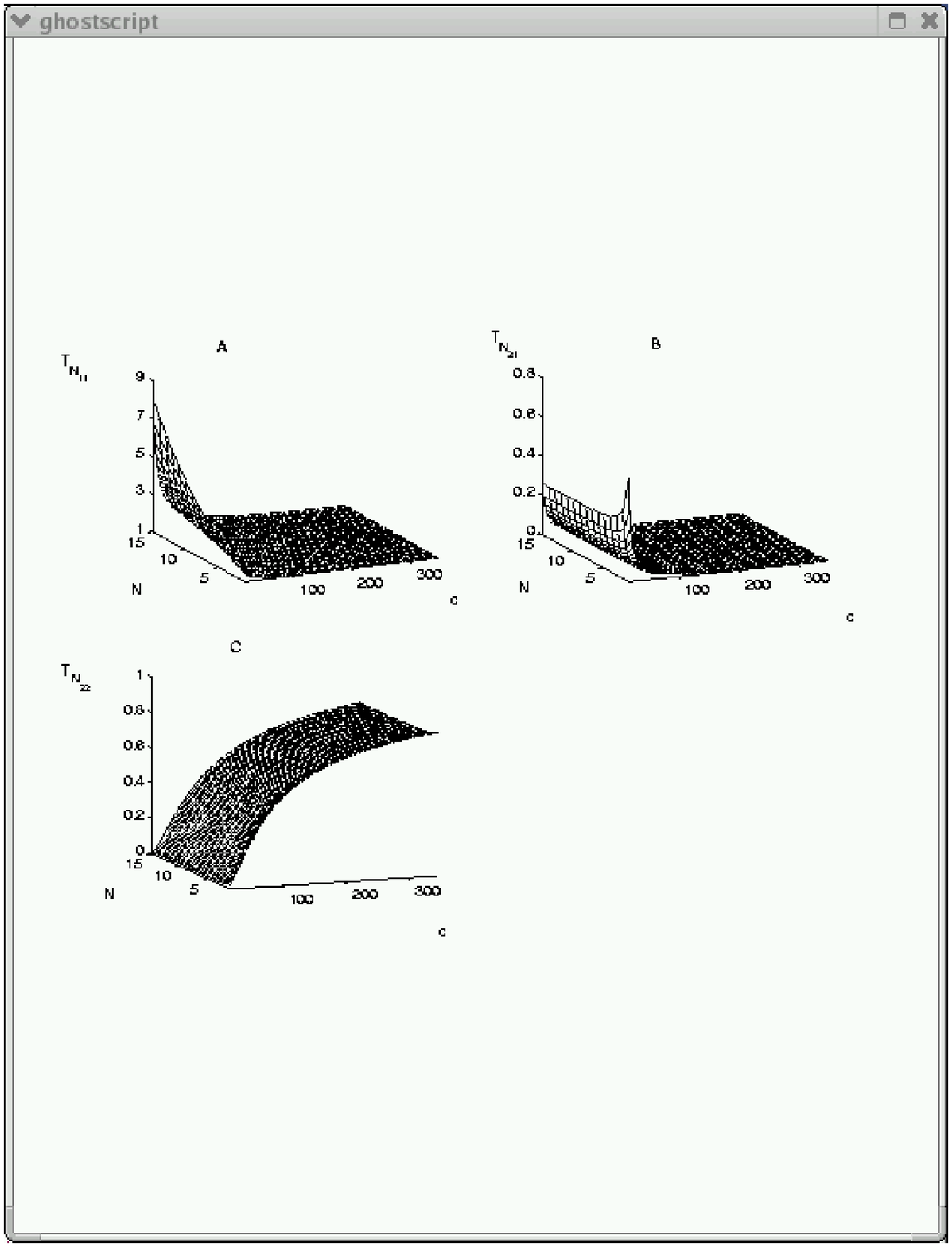}}
\caption{\label{fig:wide1}The three Panels $A-C$ show three-dimensional surfaces 
of the components $T_{N_{11}}$,  $T_{N_{21}}$  and $T_{N_{22}}$ from Eqs 
(\ref{e17}) as function of 
the ratio $c$ and the number of traps $N$ in the ranges of $300 \ge c \ge 20$
and $15 \ge N \ge 1$. The other variables are assigned the following values;
$D_o=0.8, D_i=0.4, t=k=1, L=5$.  The parameters $c$ and $N$ are obviously
dimensionless and one may  realize from Eqs (\ref{e6})-(\ref{e17}) that 
the three components $T_{N_{11}}$,  $T_{N_{21}}$  and $T_{N_{22}}$ are also 
dimensionless.  
 As seen,   the components $T_{N_{11}}$  and $T_{N_{22}}$ tend to
unity for the larger values of $c$ and $N$ and the component $T_{N_{21}}$ tends to
zero. This signifies that all the diffusing particles pass the multitrap system 
 (see Eq (\ref{e18})). }
\end{figure}

  \begin{figure}[hb]
\centerline{
\epsfxsize=5.5in
\epsffile{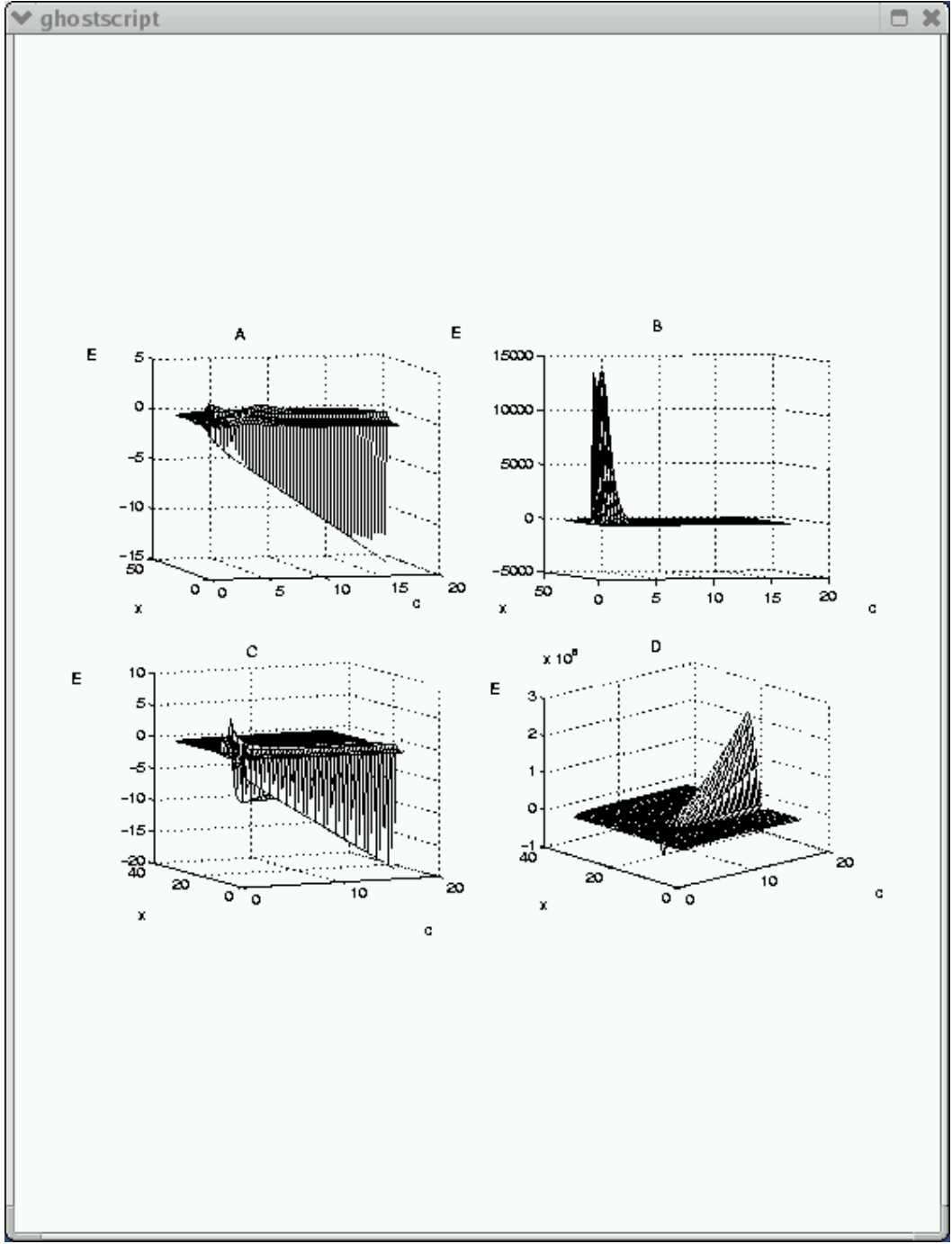}}
\caption{\label{fig:wide2}Each of the four Panels  shows 
a three-dimensional surface of the energy from Eq (\ref{e21})
as function of $x$ and the ratio $c$.  The length  $x$ is given in units of
$cm$, $c$ is dimensionless and the energy $E$ is given in units of ergs.  
   Panels $A-B$ are  both drawn for $N=2$,  $D_o=0.8$, $D_i=0.4$, $t=k=g=1$, $L=30$, 
and ${\grave x_m}=\frac{Lc}{N(1+c)}$ but $V=2$ for Panel $A$ and $V=5$ for $B$. 
 Note that
by increasing the value of $V$ from 2  to 5 results in an unexpected  
 large change  of the energy from 
$|E| \approx 10$ to $E \approx
14000$. Similar results are shown in Panels $C-D$  which are both drawn 
for  $g=k=t=1$, $L=30$, $D_o=0.8$, $D_i=0.4$,   $N=5$ and
${\grave x_m}=\frac{2Lc}{N(1+c)}+\frac{L}{N(1+c)}$  but  $V=-0.5$ for 
Panel $C$ and  $V=-0.8$ for $D$. Note the giant change 
from $|E| \approx 18$ in Panel $C$ 
to $E \approx 5\cdot 10^{8}$  in Panel $D$ that results from
 slightly  changing  from $V=-0.5$ to $V=-0.8$.  
  The negative values of $E$ result from the negative
potential energy.}
\end{figure}

\begin{figure}[hb]
\centerline{
\epsfxsize=5.5in
\epsffile{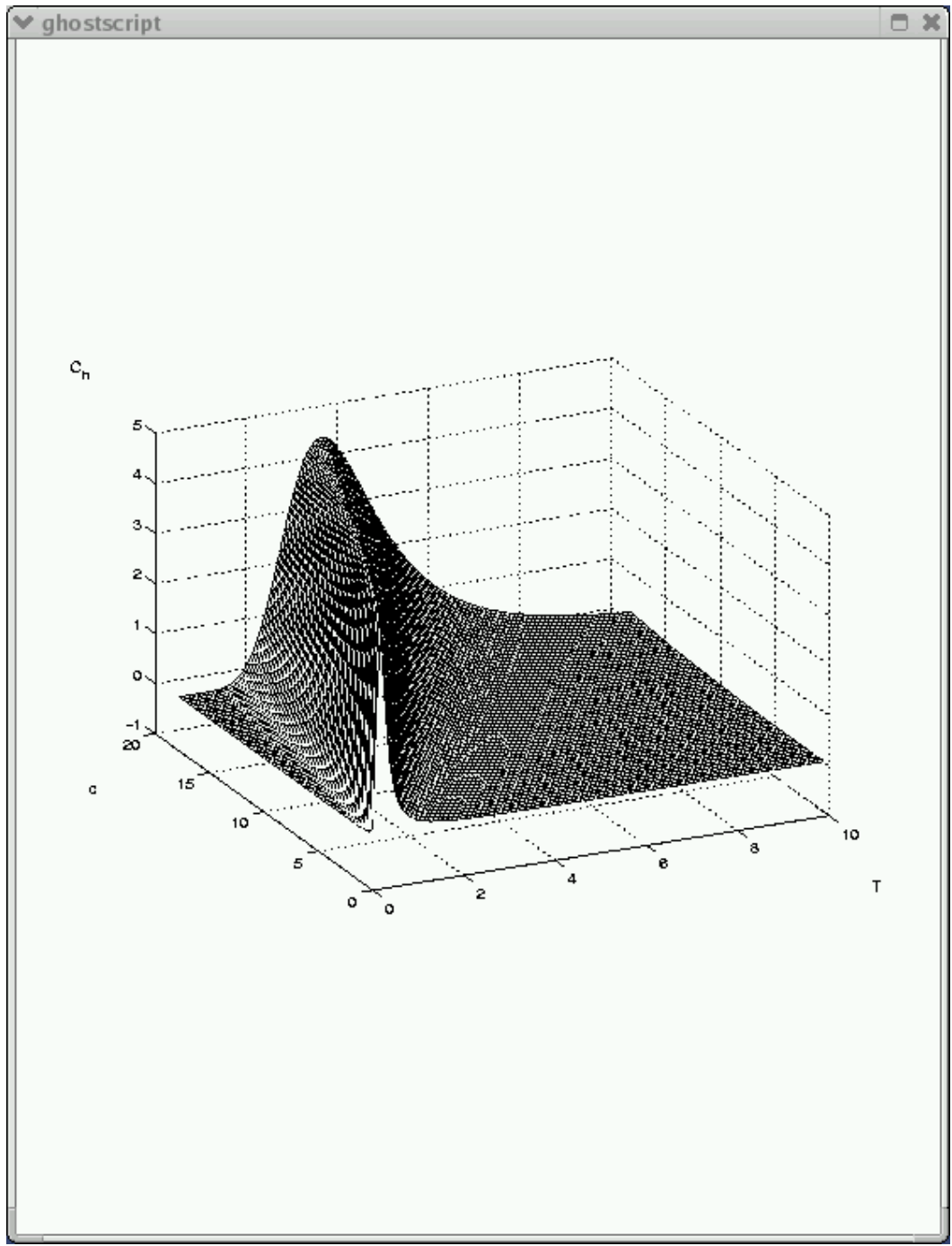}}
\caption{\label{fig:epsart}A three-dimensional surface of the specific heat $C_h$ from 
Eq (\ref{e23}) as function of the ratio $c$ and the temerature $T$ and 
in the absence of an  external field.  The units of $C_h$ and $T$ 
in this figure and
in Figures 4-6 are $[\frac{erg}{degree}]$ and  Kelvin degree respectively and
$c$ is dimensionless.  The appropriate
expression for the energy substituted  in Eq (\ref{e23}) is that from Eq
(\ref{e22}).  The  figure is drawn for  
$N=2$, $D_o=0.8$, $D_i=0.4$, $L=30$, $k=g=t=1$ and ${\grave
x_m}=\frac{Lc}{N(1+c)}$. Note the sharp edge of the surface for small 
$c$ and $T$ which becomes widened and flattened as $c$ increases. }
\end{figure}

 \begin{figure}[hb]
\centerline{
\epsfxsize=5.5in
\epsffile{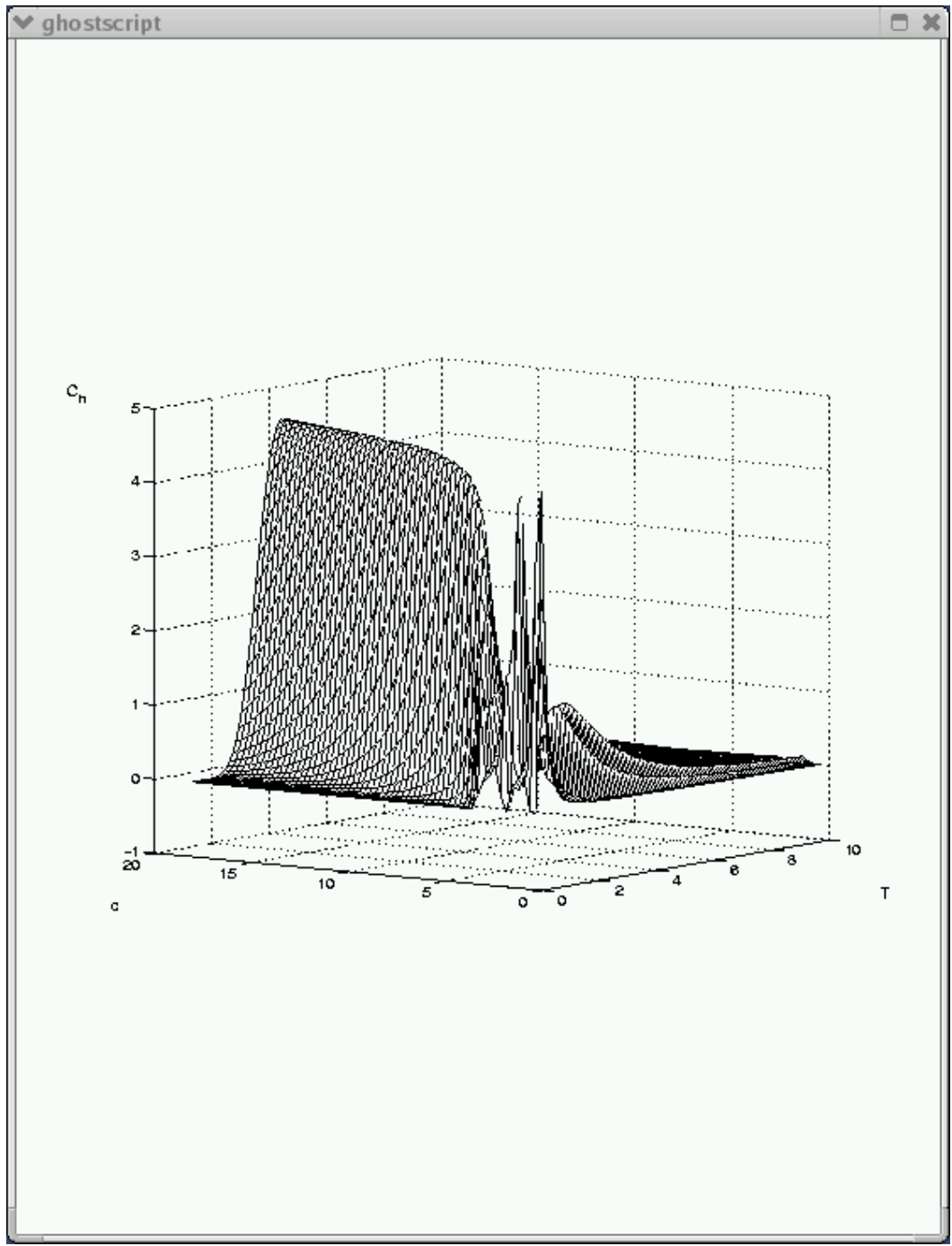}}
\caption{\label{fig:epsart2}The  specific heat $C_h$ from 
Eq (\ref{e23}) as function of $c$ and  $T$. This Figure is drawn for $V=5$, 
 $N=2$, $D_o=0.8$, $D_i=0.4$, $L=30$, $k=g=t=1$ and ${\grave
x_m}=\frac{Lc}{N(1+c)}$.
 Note that these are the values for which   Panel $B$ 
 of Figure 
 2 is drawn. From that Panel   
 we see that  changing the
value of $V$ fron 2  to 5 results in a large change of the
energy. This change is demonstrated   in the discontinuity 
of $C_h$ for either small $c$ as in here  or for larger values
of it as in Figure 5.  The units of $C_h$  and $T$ are, as remarked, 
$[\frac{erg}{degree}]$ and  Kelvin degree respectively and $c$ is
dimensionless.} 
\end{figure}

\begin{figure}[hb]
\centerline{
\epsfxsize=6in
\epsffile{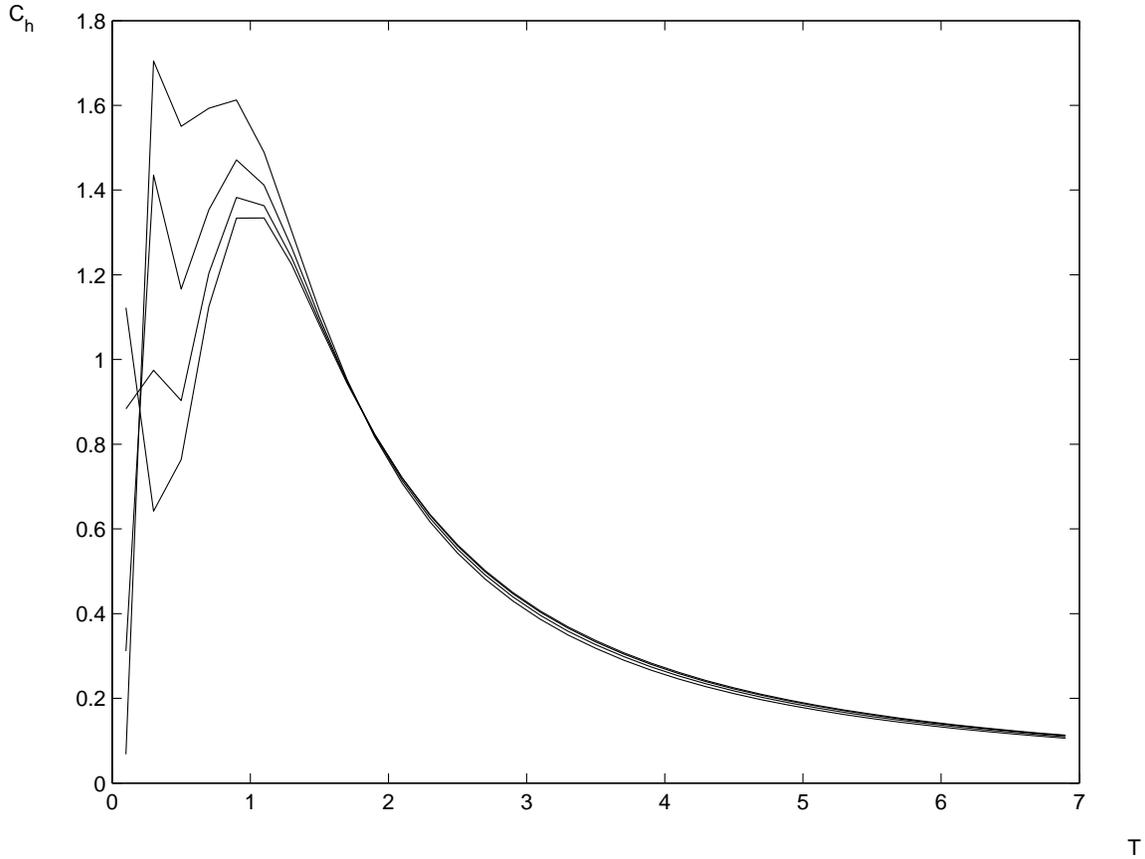}}
\caption{\label{fig:epsart3}The figure shows four different curves of the specific heat
$C_h$ in units of $[\frac{erg}{degree}]$ as function of the temperature 
$T$ in units of Kelvin for $N=2$, $V=5$, $D_o=0.8$, $D_i=0.4$,
$g=k=t=1$, $L=30$, ${\grave x_m}=\frac{Lc}{N(1+c)}$  and for 
the following 4 values of $c$; $c=3, 3.1,
3.2, 3.3$.  The double peaks are clearly seen in each curve. 
Note  that the spiky forms of Figure 4
are obtained  for exactly the same values as in this figure except 
that  $c$  is smaller. Thus, one may
conclude from  Figures 4-5 that for positive $V$ there exist  single peaks for 
small  
$c$ and double peaks for the larger values of it.} 
\end{figure}

 \begin{figure}[hb]
\centerline{
\epsfxsize=6.8in
\epsffile{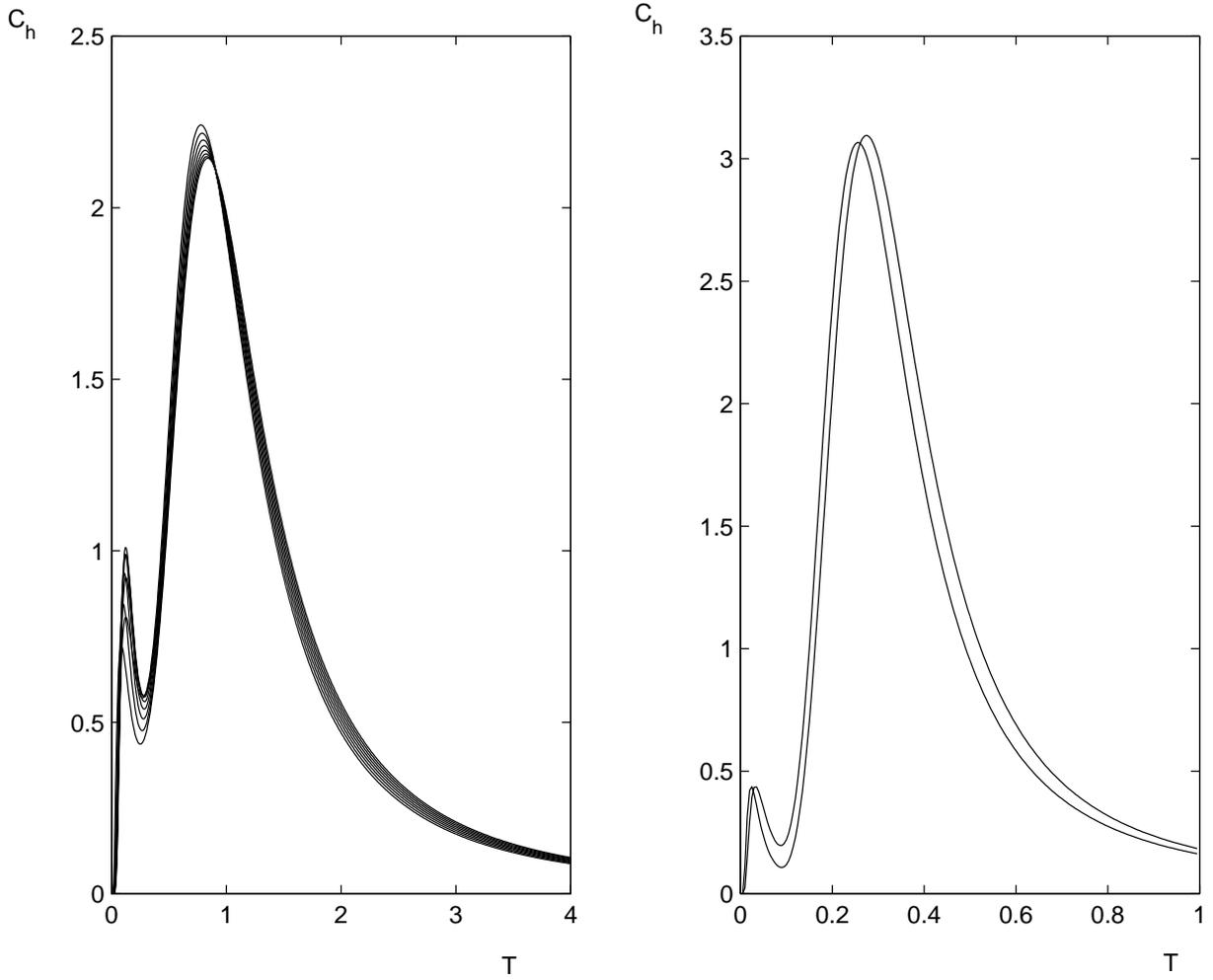}}
\caption{\label{fig:wide3}The two  Panels  show   double 
peaks of   
the
specific heat $C_h$ as function of $T$ for negative $V$ and  
$D_o=0.8$, $D_i=0.4$, $g=k=t=1$, and $L=30$.  
The Panel at the left  shows eight double-peak curves for $N=5$, 
$V=-0.37$,
${\grave x_m}=\frac{2Lc}{N(1+c)}+\frac{L}{N(1+c)}$ and the 8 values 
of $c=1.5+0.065\cdot n, \ \ n=1, 2, \ldots 7$.  The Panel at the right 
shows two double-peak curves for $N=4$, $V=-1.92$, 
${\grave x_m}=\frac{Lc}{N(1+c)}$ and
the two values of $c=0.39, 0.4$.   $C_h$ and $T$ are given in units of
$[\frac{erg}{degree}]$ and  kelvin degree respectively.  }
\end{figure}

\end{document}